\documentclass[useAMS]{mn2e}
\usepackage{graphicx}
\usepackage{amsmath}

\newcommand{\ggs}{Garc\'{\i}a--Segura\ }
\newcommand{\gd}{Garc\'{\i}a--D\'{\i}az\ }

\def\vhel{\ifmmode{V_{{\rm HEL}}}\else{$V_{{\rm HEL}}$}\fi}
\def\vsys{\ifmmode{V_{\rm sys}}\else{$V_{\rm sys}$}\fi}
\def\kms{\ifmmode{~{\rm km\,s}^{-1}}\else{~km~s$^{-1}$}\fi}
\def\vlsr{\ifmmode{v_{\rm lsr}}\else{$v_{\rm lsr}$}\fi}

\newcommand{\msolpyr}{M$_{\odot} {\rm yr}^{-1}$}
\title{A wind-shell interaction model for multipolar planetary nebulae }

\author[W. Steffen et al.]
{\parbox{\textwidth}{W. Steffen$^{1,4}$\thanks{e-mail:\texttt{wsteffen@astro.unam.mx}}
N. Koning$^{2}$,  A. Esquivel$^{3}$, G. Garc\'{\i}a-Segura$^{1}$, Ma. T. Garc\'{\i}a--D\'{\i}az$^{1}$,
J.A. L\'opez$^{1}$, M. Magnor$^{4,5}$}\vspace{0.4cm} \\
$^{1}$Instituto de Astronom\'{\i}a, Universidad Nacional Aut\'onoma de
M\'exico,Apdo Postal 877, Ensenada 22800, Baja California, M\'exico\\
$^{2}$Department of Physics and Astronomy, University of Calgary, Calgary, Canada \\
$^{3}$Instituto de Ciencias Nucleares, Universidad Nacional Aut\'onoma de M\'exico, M\'exico, D.F., M\'exico \\
$^{4}$Institut f\"ur Computergraphik, Technische Universit\"at Braunschweig, Braunschweig, Germany \\
$^{5}$Dept. Physics and Astronomy, University of New Mexico, USA
 }

\begin{document}

\date{Received **insert**; Accepted **insert**}

\pagerange{\pageref{firstpage}--\pageref{lastpage}}

\maketitle
\label{firstpage}

\begin{abstract}
\noindent
 We explore the formation of multipolar structures in planetary and pre--planetary nebulae from the interaction of a fast post-AGB wind with a highly inhomogeneous and filamentary shell structure assumed to form during the final phase of the high density wind. The simulations were performed with a new hydrodynamics code integrated in the interactive framework of the astrophysical modeling package SHAPE. In contrast to conventional astrophysical hydrodynamics software, the new code does not require any programming intervention by the user for setting up or controlling the code. Visualization and analysis of the simulation data has been done in SHAPE without external software. The key conclusion from the simulations is that secondary lobes in planetary nebulae, such as Hubble~5 and K3--17, can be formed through the interaction of a fast low-density wind with a complex high density environment, such as a filamentary circumstellar shell. The more complicated alternative explanation of intermittent collimated outflows that change direction, in many cases may therefore not be necessary. We consider that the wind-shell interaction scenario is more likely since the bow-shock shape expected from a strongly cooling bow-shock from jets is different from that of the observed bubbles. Furthermore, the timescales of the wind-wind interaction suggest that the progenitor star was rather massive.

\end{abstract}


\section{Introduction}
\label{introduction.sec}

\begin{figure}
\includegraphics[width=85mm]{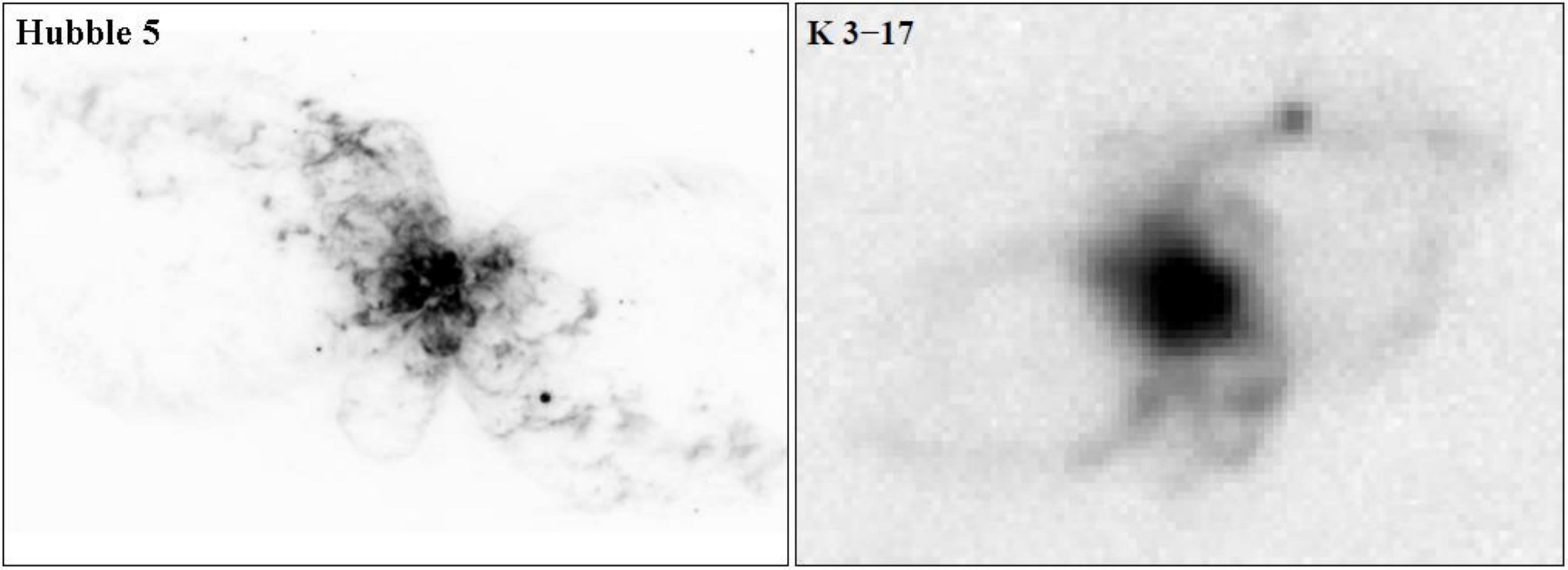}
\caption{[NII] images of Hubble~5 (HST WFPC2 archive, left) and K3--17 (MES, San Pedro M\'artir 2.1m--Telescope).  }
\label{observations.fig}
\end{figure}

Some planetary nebulae and especially pre-planetary nebulae have been found with complex multipolar structures that do not fit into the spherical interacting wind scenario (Kwok, Purton \& Fitzgerald, 1978) or its generalized version that incorporates the bipolar structures by introducing equatorial density enhancements (Kahn \& West, 1985). {\bf Especially Hubble~5 and K3-17 are peculiar, since they have large bipolar and clearly distinct multiple smaller secondary lobes (Figure \ref{observations.fig})}.

A number of possible scenarios for multipolar structures in planetary and pre--planetary nebulae have been proposed. They can be divided into three main categories. First, those with episodic ejections of collimated bipolar jets in changing directions (e.g. L\'opez, Meaburn \& Parker, 1993; Sahai \& Trauger, 1998). Second, the effects of illumination through density holes (Kwok, 2010; Koning, Kwok \& Steffen, 2013) or dynamical effects caused by illumination (\ggs, 2010) and, third, differential expansion of an uncollimated wind into a circumstellar density distribution with holes or cavities (this work).

The first scenario produces a clear signature of point--symmetry, which often can be easily identified. Likely examples for such objects are KjPn8 (L\'opez et al., 1995) or NGC~2440 (L\'{o}pez et al., 1998). Manchado, Stanghellini \& Guerrero (2006) found several similar objects in a survey and introduced the classification of ``quadrupolar" for them. Even higher multiplicity of the lobes is found in the so--called ``starfish" nebulae, such as He~2--47 and M1--37 and similar objects. They have also been attributed to the ejection of collimated jets in different directions (Sahai, 2000). Some of them (e.g. He~2--47, IRAS 19024+0044) might have multiple lobes that appear to come in pairs, which supports such a mechanism, but others, such as M1--37 or CPD-56°8032, present much more unevenly sized lobes (Chesneau et al., 2006). Hubble 5 and K3--17 have an overall bipolar structure, but present multiple secondary lobes emerging from the core and appear in apparently random directions, which lead L\'opez et al. (2012) to suggest that the secondary lobes were formed when the dense, thick core broke up.

The formation of bipolar planetary nebulae has been studied in detail by many authors (see e.g. the review by Balick \& Frank, 2002, and references therein). The formation of multipolar PNe and PPNe has been studied as a consequence of the ejection of collimated jets by Vel\'azquez et al. (2012). However, to our knowledge the dynamical formation of secondary lobes as a result of a circumstellar environment with complex structure has not been addressed numerically before. We will therefore concentrate on the mechanism to form secondary lobes in the presence of large--scale bipolar lobes.

It is not obvious whether the secondary lobes can be produced in the presence of large--scale bipolar lobes, because the latter might function as ``exhausts" or pressure drains preventing the expansion of the secondary lobes. Furthermore, the main lobes of Hubble~5 show signs of point--symmetry, hinting at a mixed case, where a collimated flow might be producing the main lobes at an oblique angle with a pre--existing equatorial density enhancement or a short collimated ejection is interacting with pre\-existing bipolar lobes. We therefore perform numerical experiments to test these scenarios for the formation of secondary lobes and try to set limits on the conditions that are necessary for them to form.

The origin of the complex structure could either be instabilities in shells in the AGB wind, e.g. due to dissociation fronts in low--mass central stars (\ggs, 2010) or thin--shell instabilities in expanding wind shells (Vishniac, 1983). Potentially, the turbulent structure of the AGB atmosphere itself, such as those found in simulations by Freytag \& H\"ofner (2008) might find their signature in the circumstellar envelope, although the timescales of a few years for the turbulence is much shorter than the envelope formation (hundreds of years). Observations of detached AGB shells show that many are indeed clumpy and filamentary (Cox et al., 2012). It is therefore reasonable to assume that any dense spherical shell will be filamentary at smaller radii, since this implies shorter time scales and therefore a stronger imprint of instabilities and atmospheric turbulence in the shells. The interaction of a low density stellar wind or thermal high pressure bubble  with the filamentary shell will then necessarily produce a complex structure.

Spatially resolved detached shells around AGB stars have been found at distances of the order of 0.1~parcsec (Izumiura, et al., 1996; Cox et al., 2012), which is of the same order as the size of pre-planetary nebulae. The size of the region from which the secondary lobes in Hubble~5 emerge, appear to be of the same order of magnitude (assuming a rather uncertain distance of more than 3~kpc, Pottasch \& Surendiranath, 2007; L\'opez et al., 2012). In this paper we therefore explore the formation of multipolar PNe and PPNe from the interaction of a fast low--density wind or high pressure bubble with a thin spherical shell of uneven density distribution that has a radius of a few times $10^{14}$~meters.

The particular questions that we address in this paper are: Can multipolar structures be generated by a spherical fast wind that is interacting with a highly clumpy or filamentary dense shell-like structure around the central star? Can such a model produce secondary lobes in the presence of a dominant bipolar structure similar to those in the planetary nebulae Hubble~5 and K3--17?

\section{Interactive hydrodynamics in Shape}
\label{shape.sec}

The hydrodynamical simulations have been done with a new hydrodynamic code that has been incorporated in the SHAPE 3D modeling and visualization environment. We present our prototypical hydrodynamics module which takes the initial and boundary conditions from the 3--D environment. The spatial distribution of the initial physical conditions, such as density, temperature and velocity, is set up interactively and does not require any programming by the user. {\bf More information on SHAPE and the hydrodyanmics module, including code test simulations, can be found on the ``Learning Center" page of the {\em ShapeSite} at {\em http://www.astrosen.unam.mx/shape}}.

SHAPE is an astrophysical modeling software package that has extensively been used for 3--D morpho--kinematic modeling of planetary nebulae (e.g. \gd et al., 2012; Chong et al., 2012; Santander--Garc\'{\i}a et al, 2012) and similar astrophysical objects such as nebulae around novae (Ribeiro et al., 2011). The model geometry, velocity field and parameters for radiation transfer are set up interactively in a similar fashion as is common in computer aided design (CAD) or engineering software (Steffen et al., 2006, 2011) using polygon mesh objects that are filled with physical properties as functions of position. These functions of position can either be given as analytic functions or manually set Bezier curves. These can be based on external data or points that have been introduced interactively via the graphics interface. Similarly time variability of parameters is introduced through similar functions in a special animation module, that also allows control of output as a function of time.

\begin{figure}
\includegraphics[width=85mm]{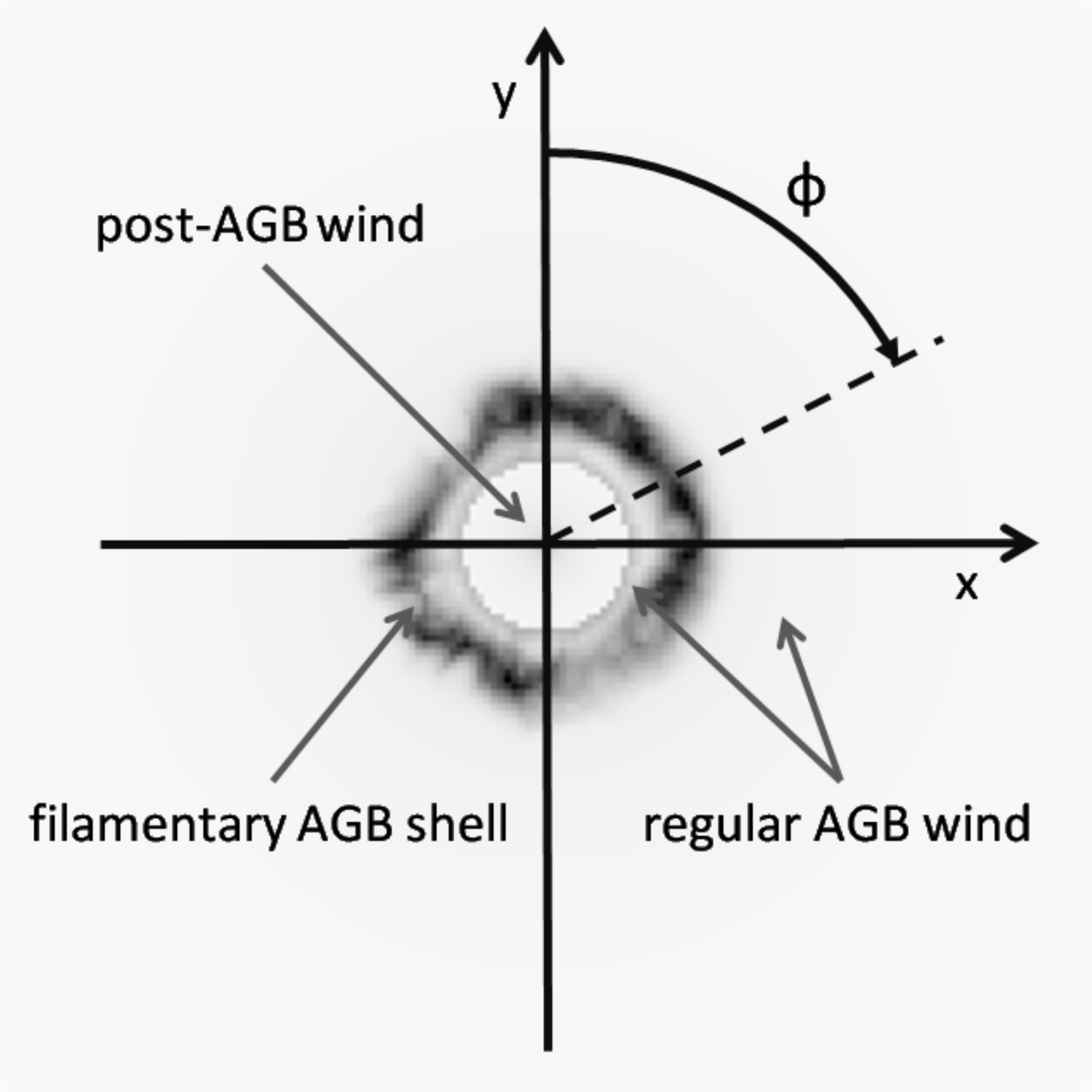}
\caption{Example of initial conditions and coordinate setup for the description of pole--equator asymmetries of the flow parameters. The equator corresponds to the xz-plane. The grey-scale is a cut through the density distribution. The labels indicate the main flow regions. Note that only the high density shell has been initialized with the turbulent structure shown in Figure (\ref{noise.fig}). }
\label{sketch.fig}
\end{figure}

\subsection{Numerical scheme}
\label{numerical.sec}
In an effort to take this interactive simulation framework a step further, we have incorporated a hydrodynamical module that solves the basic inviscid compressible Euler equations for the conservation of mass, momentum and energy as given in Equation (\ref{eq:euler}) on a regular grid.
 \begin{equation}
\frac{\partial U}{\partial t} + \frac{\partial F}{\partial x} + \frac{\partial G}{\partial y} + \frac{\partial H}{\partial z} = S
\label{eq:euler}
\end{equation}
In these equations $U$ is a vector with the so-called conserved quantities and $F$, $G$ and $H$ are the conserved fluxes along the cartesian coordinate directions $x$, $y$ and $z$, respectively, as given in Equations (\ref{eq:U} -- \ref{eq:H}). In the latter equations $u$, $v$, and $w$ are the velocity components along the coordinate directions, respectively. $S$ represents sources and sinks of the ``conserved" quantities and $t$ denotes time.
 \begin{equation}
U = [\rho, \rho u, \rho v, \rho w, E]
\label{eq:U}
\end{equation}
 \begin{equation}
F = [\rho u, \rho u^2+P, \rho uv, \rho uw, u(E+P)]
\label{eq:F}
\end{equation}
 \begin{equation}
G = [\rho v, \rho uv, \rho v^2+P, \rho vw, v(E+P)]
\label{eq:G}
\end{equation}
 \begin{equation}
H = [\rho w, \rho uw, \rho uw, \rho w^2+P, w(E+P)]
\label{eq:H}
\end{equation}
The set of Euler equations is closed by the equation of state of for an ideal gas. Cooling is included via a parameterized cooling curve as a function of temperature and density. 
 {\bf Since in this study we are interested in the qualitative aspects of the formation of multi-polar structures, rather than in a quantitative comparison of physical properties with observed objects, we have introduced an efficient simplified treatment of radiative cooling. It is a simple analytic function of temperature roughly approximating the cooling curve by Dalgarno \& McCray (1972) by three straight lines in logarithmic space of temperature and energy loss. The initial footpoint is at $10^4$~K, below which we assume the cooling to be negligible, i.e. photo-ionization equilibrium is assumed once the gas has been shock-heated. The peak is at $10^{5.5}$~K and free-free emission takes over at temperatures higher than $10^{7.5}$~K. The final energy loss per unit volume is then computed from $S_{\rm cool} = n^2 \Lambda(T)$, where $\Lambda$ is the cooling curve, $n$ is the number density of particles (assumed fully ionized) and $T$ is the temperature. }
 
The numerical scheme follows that of Raga, Navarro-Gonz\'alez \& Villagrain-Muniz (2000) except that it currently runs on a regular grid without adaptive mesh refinement.
{\bf The method is a finite volume upwind shock-capturing Godunov scheme. The intercell fluxes are calculated with the flux--vector splitting of van Leer (1982). To achieve second order accuracy we do linear reconstruction of the primitive variables with slope limiters (van Leer 1974). Although the method is somewhat diffusive, we need to introduce an artificial viscosity to control the numerical artifacts that otherwise occur due to strong shocks propagating aligned with the coordinate axes. The viscous term is of the form $\eta \nabla^2U$, with a value of $\eta=10^{-4}$. We used a CFL (Courant, Friedrichs \& Lewy 1967) number of 0.4 in all the models.}

The code was implemented in object-oriented Java and was parallelized using threading for multi-core CPUs. Real--time interactive pre-visualization allows the user to follow the simulation as it progresses and at the same time rotate and fly-through the 3--D volume rendering of density, pressure, temperature and velocity. More detailed analysis and visualization, such as spectral line images, P--V diagrams as well as other diagnostic graphs can be displayed in other SHAPE modules at any time.

\begin{figure}
\includegraphics[width=85mm]{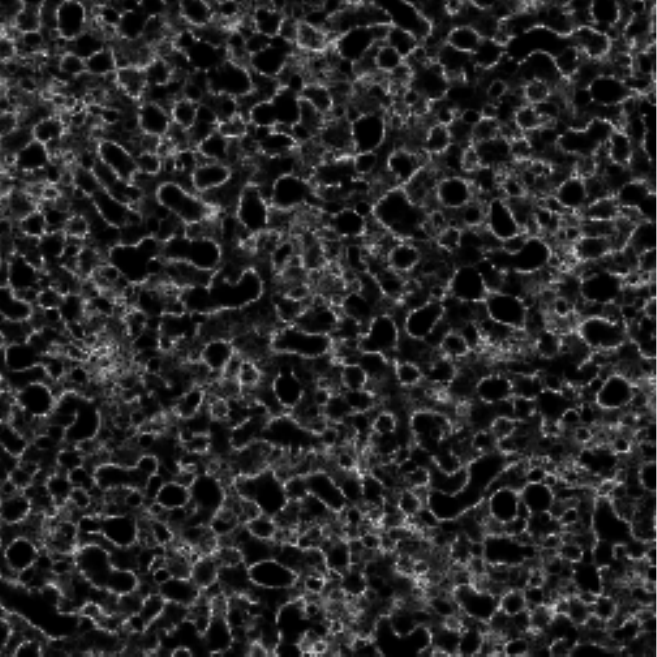}
\caption{An example cut through the 3D turbulent noise structure that has been used for the dense shell. This pattern would fill the complete computational domain, but the shell uses only the section that fills its volume enclosed by a polygon mesh. The main parameters for the filaments are the size scale of the structures and an exponent that controls the density contrast between the filaments and the voids. Since all structures are quasi-random, the exact structure and location of features is not determined by the user. However, the complete pattern can be shifted along the coordinate axes by any amount.  }
\label{noise.fig}
\end{figure}

\subsection{Interactive setup and boundary conditions}
\label{setup.sec}

In computer aided design or computer animation for the media industry 3--D applications heavily rely on interactive modeling and real-time (pre-)visualization of 3--D objects and scenes. Often they also include complex physical computations, including particle and fluid dynamics. Fluid dynamical simulations can interact with the ambient objects in very complex ways. The setup of such simulations is done in an interactive interface, which allows quick pre-visualization of the progressing simulation. This allows the user to
easily change the initial setup or apply modifications while the simulation is running. Furthermore, complex time variability can be set up within the interactive animation modules, allowing a graphical representation and manipulation of simulation parameters as a function of time.

In SHAPE we have adopted a similar approach, where all necessary parameters and elements in the environment
are set up in an interactive fashion while still allowing an algebraic description via a symbolic math interpreter (Steffen et al., 2011). Initial ``primitive" mesh objects such as spheres, cylinders or cubes are added to the scene and then modified and positioned to yield a more complex structure. The mesh represents the bounding volume of an object and may be filled or used as a shell. A distribution of physical parameters, such as density and velocity, is then applied as a function of position and time.

Initial and internal boundary conditions are set up as mesh objects in the 3--D modules. Any object can be used as an initial or internal boundary condition, according to the setting of a corresponding flag. For an object that is used only as initial conditions, the space it occupies is set to its properties only at the beginning of the simulations, while for an internal boundary condition, the corresponding grid cells are reset after every timestep. The boundary conditions at the edges of the computing domain in all simulations have been set to be gradient free (outflow).

\begin{figure}
\includegraphics[width=80mm]{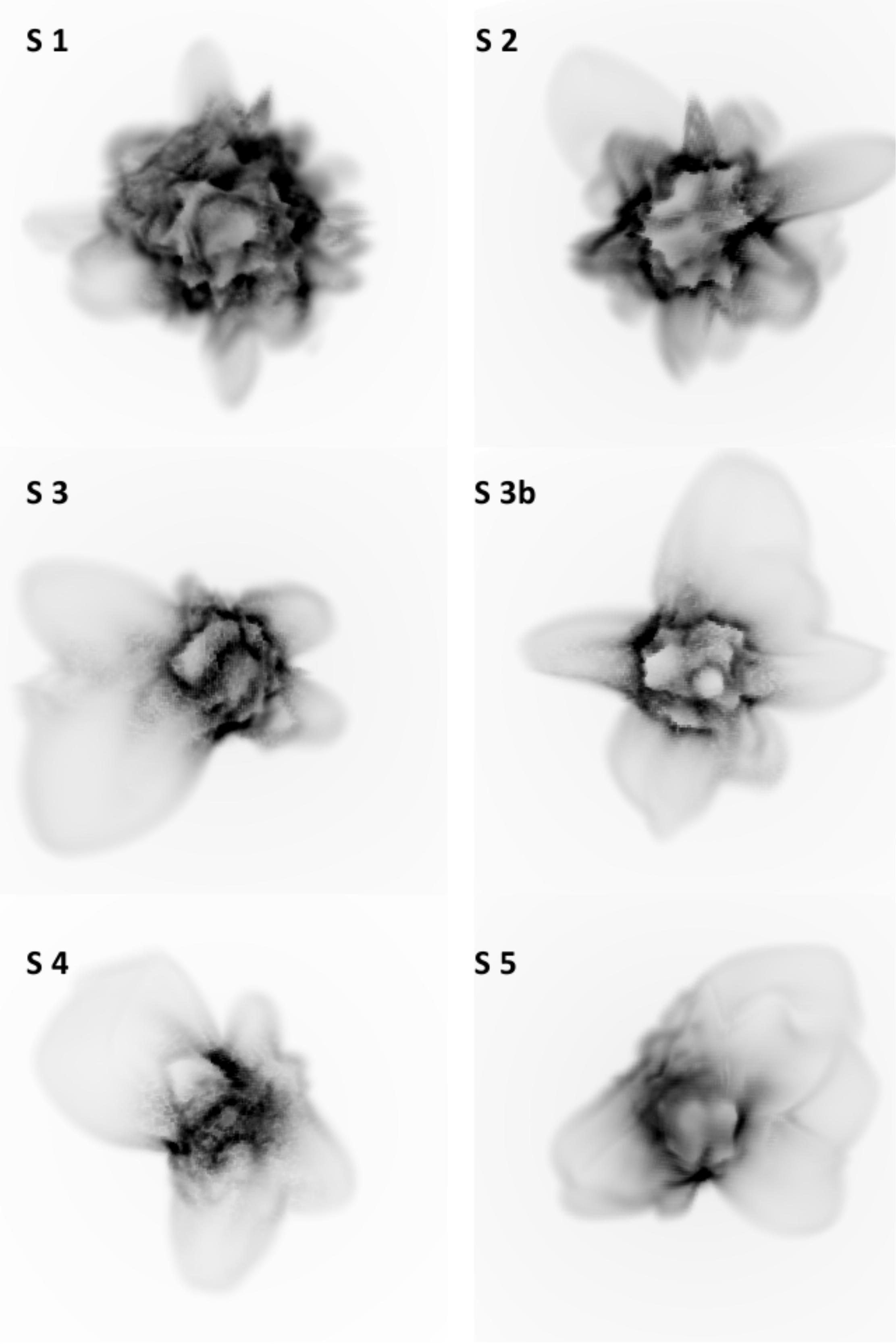}

\caption{Different starfish nebulae (Models S1 to S5) generated with increasing size scales of the noise patterns in the dense shell. The pattern scale increases from the top left to the bottom right. The middle row shows two different realizations with the same size scale for the noise pattern. All other parameters are identical among the six simulation. The inverted image intensity corresponds to the integral of the emission measure along the line of sight, with a power-law gray-scale with an exponent of 0.25.}
\label{starfish.fig}
\end{figure}

\begin{figure}
\includegraphics[width=40mm]{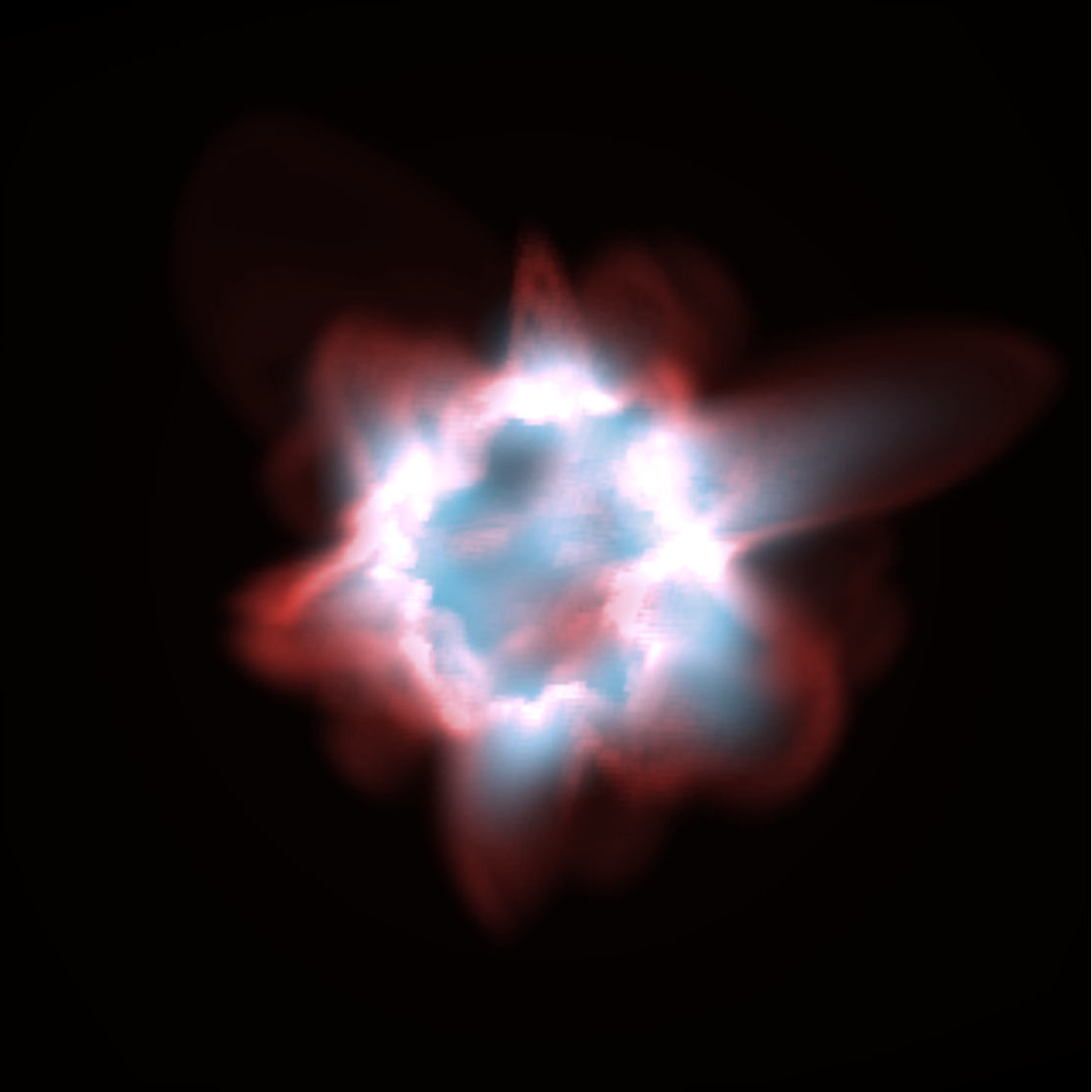}
\includegraphics[width=40mm]{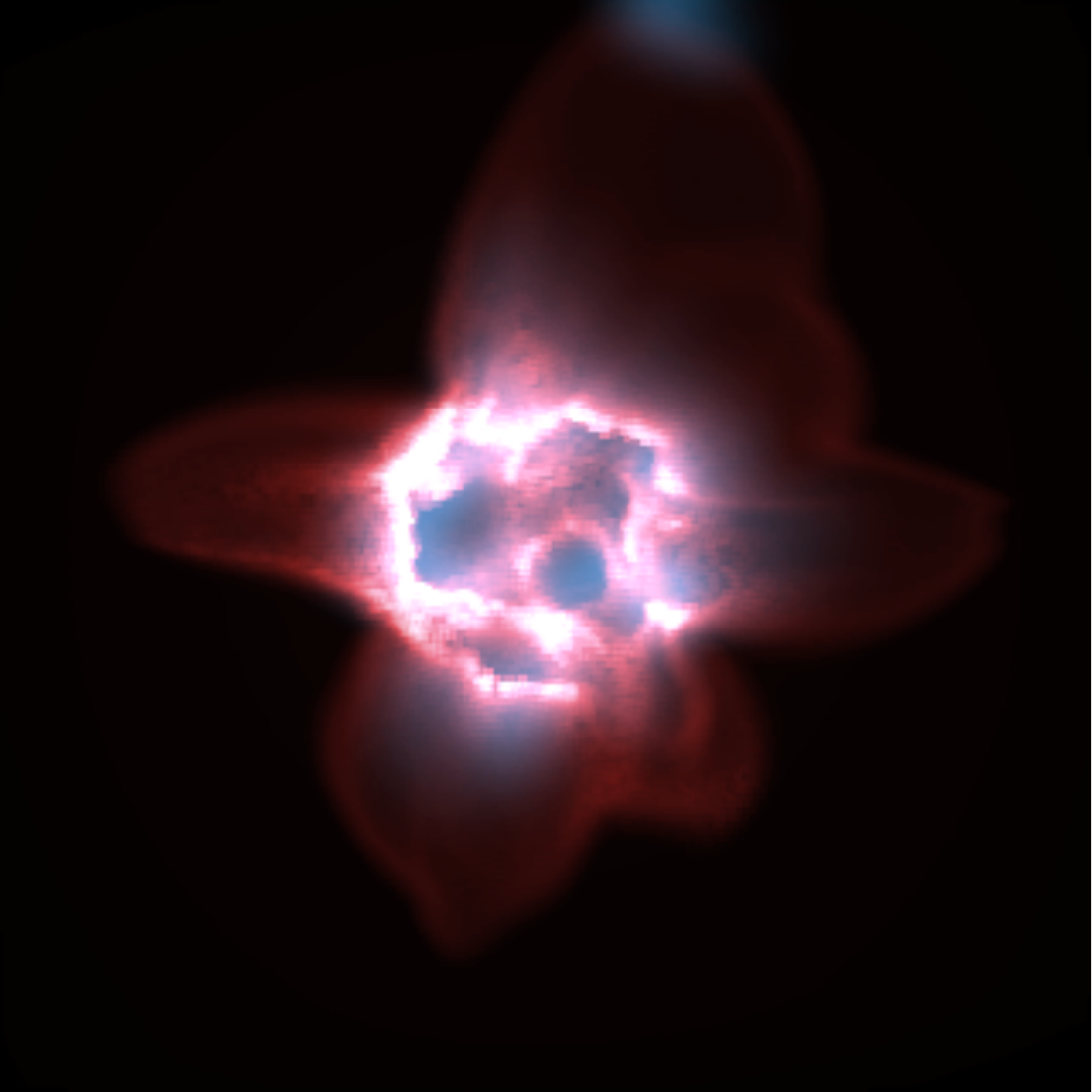}

\caption{Rendering of models S2 (left) and S3b with the emission measure from cool dense gas color coded in red and free-free emission from gas with a temperature higher than $10^6$~K coded in blue. A color version of this figure is available in the electronic version of the paper.}
\label{starfish_xray.fig}
\end{figure}

\section{Hydrodynamical models}
\label{models.sec}

We consider the interaction of a constant tenuous fast wind of mass--loss rate $\dot M_{\rm w}$ and velocity $v_{\rm w}$. This wind interacts with an environment that has been created from a variable AGB--wind of $\dot M_{\rm agb} (t,\theta, \phi)$, where $t$ is time and $\theta$ and $\phi$ are the azimuthal and polar angles on a sphere around the central star, respectively. {\bf Figure (\ref{sketch.fig}) shows a schematic description of our setup.} Typically the expansion velocity of the AGB--wind is of order 15 \kms, while the fast wind can have velocities up to the order of 2000 \kms (S\'anchez Contreras \& Sahai, 2001). Considering the limited spatial resolution and size ratio between the smallest and largest structures in the simulations, we want to reduce the expansion of the dense shell with respect to the lobes as much as possible. Therefore the velocity of the ambient medium is initially set to zero. The density of the ambient medium, and in particular that of the dense filamentary shell, is chosen such that the speed along the line of sight of the bipolar lobes roughly corresponds to that observed in Hubble~5, which is of order 200 \kms (L\'opez et al.,2012).

The key hypothesis for this paper is that the shock produced by the wind interaction propagates at speeds that are highly dependent on the direction due to the changing density gradient between the AGB-- and the fast wind. Furthermore, we assume that the main interaction occurs with a filamentary thick high density shell with a local density $\rho_(\vec{r})$.  Therefore, let us consider the expansion speed of a wind shock as a function of locally changing AGB wind density $\rho(\vec{r})$. The expansion speed $v_{\rm s}$ of the shock at a distance $r$ is then
\begin{equation}
v_{\rm s} = v_{\rm w} \left(\frac{\zeta \rho_{\rm w}}{\rho_{(\vec{r})} } \right)^{1/2} = \frac{1}{r} \left( \frac{\zeta \dot{M}_{\rm w} v_{\rm w}}{4\pi \rho_{(\vec{r})}} \right)^{1/2}
\label{shockvelocity.eq}
\end{equation}
where $\zeta = 1$ and $\zeta = 9/16$ for an isothermal and adiabatic shock, respectively (Steffen \& L\'opez, 2004). In the case of an isotropic fast wind the velocity ratio is thus determined only by the corresponding density ratio. The ratio of expansion speed at two different positions $\vec{r}_{\rm 1}$ and $\vec{r}_{\rm 2}$ will then be
\begin{equation}
\frac{v_{(\vec{r}_{\rm 1})} }{v_{(\vec{r}_{\rm 2})} } = \left( \frac{ \rho_{(\vec{r}_{\rm 1})} } { \rho_{(\vec{r}_{\rm 2})} } \right)^{1/2}
\label{velratio.eq}
\end{equation}
Here we assumed that the $\zeta$ parameter is the same for both shocks.

Kinematic observations of the secondary lobes and the inner high density regions of Hubble~5 (L\'opez et al., 2012) show that the expansion speed of the secondary lobes is around 100\kms, where as the bright high density central region probably expands at less than 30\kms. Subtracting a typical 15\kms velocity from the original AGB wind speed, from Equation (\ref{velratio.eq}) it follows that the density ratio between the voids where the bubbles escape and the constraining filaments should be at least of order 50.

Similarly, if the velocity of the fast wind is around 1000 \kms (appropriate for a high mass central star like the one expected in Hubble~5, Pottasch \& Surendiranath, 2007), and with a typical expansion speed of a bubble of order 100\kms, then the density ratio between the wind and the voids has to be of order 100 or more. Hence, combining the density contrast between filaments and voids, the ratio between the filaments and the fast wind has to be of order 5000. Most of the shell mass is therefore concentrated in the filaments and will remain near the central star rather than expand with the bubbles. The extreme density contrast between the inner and outer regions explains the very high brightness contrast between the inner region and the bubbles in objects such as Hubble~5 and K3-17.

The hydrodynamic models were set up in the 3D--module of SHAPE with two different spherical mesh objects: the dense AGB wind, assumed static for all simulations, and the low density fast wind imposed at every time step within a spherical region of radius $2\times10^{14}\ {\rm m}$. The cubic computational domain has a physical width of $1.0\times10^{15}\ {\rm m}$. The AGB wind setting is applied as an initial condition.

The density $\rho_{\rm w}$ within the internal boundary condition for the fast wind is set such that it provides an ``equivalent" mass--loss rate of $\dot M$ at the equator. In the models where the wind density and velocity vary with the angle from the equator $\theta$, we have chosen interpolation functions between the equator and poles that allow control about how the density and velocity smoothly change with direction. Note that the direction of gas motion is kept radial. The density and velocity are then given by the following equations:
\begin{equation}
\rho_{\rm w}(\theta) = \frac{\dot M}{4\pi v_{\rm w0} r^2} \left(1+(a_{\rho}-1) \frac{2|\theta|}{\pi}\right)^{b_{\rho}}
\label{rho_wind.eq}
\end{equation}
\begin{equation}
v_{\rm w}(\theta) = v_{\rm w0} \left(1+(a_{\rm v}-1) \frac{2|\theta|}{\pi}\right)^{b_{\rm v}}
\label{v_wind.eq}
\end{equation}
where $v_{\rm w0}$ is the wind speed at the equator and $r$ is the distance from the star, whereas $a_{\rho}$, $b_{\rho}$, $a_{\rm v}$ and $b_{\rm v}$ are constant parameters of the interpolation from the equator to the poles.

The details of the transition between the slow and fast winds as a function of time are largely unknown. However, the timescale for the transition appears to be short for massive progenitor stars, i.e. less than a hundred years as derived from the kinematics of proto-planetary nebulae (S\'anchez-Contreras \& Sahai, 2001; Bujarrabal et al., 1998; Cox et al., 2000; Trammel \& Goodrich, 2002). This is a small fraction of the kinematic age of Hubble~5 (of order 1000 years). Therefore, we make an instantaneous change from slow to fast wind and parameterize the AGB- and post--AGB winds separately. The AGB--wind is set as an initial condition and the fast wind as an internal boundary condition at the numerical injection distance $r_0$ with velocity $v(\theta)$ and density $\rho_w(\theta)$ as a function of angle $\theta$ from the equator. The direction of the velocity is assumed to be radial from the central star.
\begin{equation}
    \rho_{\rm w}(r_0,\theta) = \xi \rho_{\rm agb}
    \label{n_theta.eq}
\end{equation}
where $\xi$ is the density ratio between the AGB- and post-AGB wind at distance $r_0$. The ambient density $\rho_{\rm a}$ is
\begin{equation}
    \rho_{\rm a}(\vec r) = \rho_{\rm agb}(\vec r) + \rho_{\rm s}(\vec r)
    \label{n_theta.eq}
\end{equation}
where $\rho_{\rm s}(\vec r)$ is a density enhancement in a shell as defined below.
The dense inhomogeneous AGB-shell with density $\rho_{\rm s}$, that we postulate to be the main cause of the multipolar structure in our model, is added to the smooth AGB-wind as a shell with a Gaussian Envelope as a function of distance $r$ with peak density at $r_{\rm s} = 2\times10^{15}{\rm m}$. The density distribution is modulated with a filamentary ``turbulent" noise structure ${\rm turb}^p(x,y,z)$ as described in the next section (\ref{noise.sec}).
\begin{eqnarray}
 \rho_{\rm s}(\vec r)     &=&   \rho_{\rm agb} \cdot f_{\rm sr}(r) \cdot f_{\rm s\theta}(\theta) \cdot  {\rm turb}^p(x,y,z)\\
 {\rm where} && \nonumber \\
 f_{\rm sr}(r)                 &=&    \eta \cdot exp( - (r-r_{\rm s})^2/2c^2)  \nonumber  \\
 f_{\rm s \theta}(\theta) &=&   1 - (1-w) \cdot {\rm abs}({\rm sin}^b(\theta)) \nonumber
 \label{shell.eq}
 \end{eqnarray}
Here $\eta$ is the peak density ratio between the smooth AGB-wind and the AGB-shell, $w, b, c$ and $q$ are constants that determine the smooth density variation of the shell as a function of distance and angle from the equator.

\subsection{Filamentary structure of the AGB-shell}
\label{noise.sec}

There is evidence that the mass-loss of AGB stars is highly structured (Cox et al., 2012) in addition to any systematic density gradients from the equatorial plane to the poles. Therefore the propagation of the ensuing low-density fast wind through this local environment will depend on the direction with respect to the center of the star.  The type of structures that will emerge from the wind-wind interaction will strongly depend on the angular scales and density contrast of the random variations. The origin of complex density variations may be various, such as instabilities (Garc\'{\i}a-Segura 2011) or possibly the large-scale turbulence of the AGB stellar atmosphere (Freytag \& H\"ofner, 2008). In this paper we will not explore the origin of the density variations, rather than whether such variations can produce the observed multi-polar PPNe or PNe and what are the main properties of the density variations that produce these structures. This might shed some light on the possible mechanisms responsible for such structures.

In order to generate the inhomogeneous structure of the dense AGB-shell in our numerical experiment, we employ a {\it noise} generation technique common in computer graphics. For this work we chose to test different representative noise patterns based on Perlin noise (Perlin, 2002).  For the generation of secondary bubbles the {\it sparse convolution algorithm} by Lewis (1989)(Figure \ref{noise.fig})  turned out to be the most suitable with a pattern that is reasonably realistic in the astrophysical context when used together with the turbulence emulating function given in Equation ({\ref{turbulence.eq}).
\begin{equation}
{\rm turb}(x,y,z) = \sum_{i=0}^{k} \left| \frac{{\rm noise}(s_{\rm t} 2^i x, s_{\rm t} 2^i y,s_{\rm t} 2^i z))}{2^i} \right| / \sum_{i=0}^{k} 2^i
\label{turbulence.eq}
\end{equation}
Here $noise(x,y,z)$ is the improved Perlin procedural noise (Perlin 2002). The parameters $k$ and $s_{\rm t}$ determine the complexity and the spatial scaling of the turbulence pattern, respectively. They are controlled by the user. To obtain the turbulence pattern the background density is multiplied by the ${\rm turb}^p (x,y,z)$ function, which is valued in the interval [0,1]. The exponent $p$ controls the contrast level in the density distribution.

The sparse convolution noise can be set to show a filamentary structure that resembles that of a ``Swiss cheese" (Figure \ref{noise.fig}). Compared to other common noise generators, this algorithm has the additional advantage of showing fewer regular grid artifacts. Since it is generated on a regular cartesian grid, some grid alignment effects remain, which leads to some bubble structures that are stronger along the x,y,z axes. The true structure of the AGB wind is likely to be more compressed as a function of distance from the star, but the most important element here is the angular distribution of structure, since the fast wind will basically sweep up and flatten the structures along the radial directions. The key parameter that we vary is the density contrast between the voids and filaments and the overall scale of the structures. \\

\section{Results}
\label{results.sec}

\subsection{Starfish nebulae}
\label{starfish.sec}
First of all we explore the influence of the size scale of the filamentary structure of the dense AGB-shell
on the shape of the expanding fast wind. The starfish models have been computed with a grid of $160^3$ cells.
The smooth background AGB wind was assumed to have a mass-loss rate of $5\times 10^{-8}$ \msolpyr, whereas the peak mass-loss rate in the dense shell is formally $2500\times$ higher, resulting in $1.25\times 10^{-4}$ \msolpyr. For the given filamentary structure, the actual mass-loss rate is 2-3 times smaller, since the algorithm to generate the density structure of the shell locally scales down the density by a factor $\leq 1$. The parameters that vary among the models are listed in Table \ref{starfish.tab}. In order to maintain the ambient medium stationary during the simulation, except for the interaction with the fast stellar wind, initially the pressure has been set constant through the computational domain at $2\times10^{-12}$ Pascal.

Figure (\ref{starfish.fig}) shows how the starfish structure changes with increasing size scale of the filamentary structure in the dense shell. If the size scale is much smaller than the radius of the shell only small bubbles appear which merge and barely allow the formation of individual bubbles (Model S1). If the size of the voids becomes of similar scale as that of the shell radius, then the structure can have the appearance of almost a single irregular bubble that is highly asymmetric (Model S5). Models S3 and S3b are different realizations of the same noise statistics of the dense shell. They illustrate the variation or similarity of two models with the same general parameters, but different random filament positions.

As the size scale of the structures in the dense shell decreases, as expected, the size of the bubbles becomes smaller, its number larger and the overall shape of the nebula more spherical. In Model S1 most bubbles already begin to merge. In models with even smaller structures (not shown), most bubbles merge and hardly escape from the dense shell which is also accelerated by the fast wind. The small merging bubbles appear like an irregular low-density halo around the high density shell.

Although it is not a big surprise, it is worth noting that the smaller the size scale of the void structure of the dense shell, the slower is the expansion of the bubbles compared to the expansion of the constraining filaments. This can be clearly seen in Figure (\ref{starfish.fig}), where the ratio of the bubble size to that of the dense shell increases from Model S1 to S5. Due to the relatively stronger expansion of the shell for smaller void structures, the peak density of the shell at a given time is smaller compared to the cases with larger bubbles.

Since X-ray emission is an important diagnostic tool for the dynamics and energetics of PNe, in Figure (\ref{starfish_xray.fig}) we show images with the expected soft X-ray emission for hot gas of temperature above $10^6$~K (blue shading). The strongest emission is found from the wind interaction zone in the interior of the bubble. However, shocks inside the bubbles do also show some X-ray emission. Since much of the dense shell might contain dust and hence absorb most of the X-rays from the interior of the shell, the bubbles might be of comparable or higher X-ray brightness than the interior.

We conclude that starfish nebulae can be formed by a spherically symmetric fast stellar wind interacting with a dense filamentary shell, that might have its origin in the final stages of the AGB mass-loss period.

\begin{table}
\begin{minipage}{80mm}
\caption{The variable parameters of the {\it starfish} models.}
\label{starfish.tab}
\begin{tabular}{l*{3}{r}}
Model              &Time           &  Noise scale & \\
Unit              &$10^{10}$ sec& $10^{12}$m \\
\hline
S 1           & 2.00 & 4.0    \\
S 2           & 1.50 & 7.5     \\
S 3           & 1.00 & 10.0   \\
S 3 b        & 1.00 & 10.0   \\
S 4           & 0.85 & 15.0   \\
S 5           & 1.00 & 20.0   \\

\hline
\end{tabular}

\end{minipage}
\end{table}

\begin{table*}
\begin{minipage}{150mm}
\caption{The physical parameters that vary between the different bipolar models.}
\label{bipolars.tab}
\begin{tabular}{l*{14}{r}}
Model              &Time         & Incl.        & $v_f$ &   $\rho_{\rm agb}/\rho_{\rm w} $   & $p$  &  $w$ & $b$  &  $\eta$ & $s_{\rm t}$   & $a_{\rm \rho}$ & $b_{\rm \rho}$ & $a_{\rm v}$ & $b_{\rm v}$ & $\phi_{\rm j}$\\

Unit                  &$10^{10}$ sec& $^{\circ}$   &    \kms   &   &         &          &          &              & $10^{12}$~m &
&                          &                     &                      &    \\
\hline
B 2           & 0.70 &   0 & 600 & 30 & 45  & 0.05  & 1.3 & 1000 & 6 & 3 & 1.0 & 3 & 2.0 &   0 \\
B 6           & 0.96 & 20 & 700 & 30 & 45  & 0.02  & 0.5 & 2000 & 5 & 3 & 1.0 & 2 & 2.0 & 20 \\
B 9           & 0.70 & 20 & 800 & 30 & 50  & 0.02  & 2.0 & 2000 & 6 & 3 & 1.5 & 2 & 1.5 &   0 \\
B 14         & 1.20 &   0 & 500 & 30 & 45  & 0.02  & 2.0 & 2000 & 6 & 3 & 1.5 & 2 & 1.5 & 30 \\
B 15         & 1.60 &   0 & 700 & 30 & 25  & 0.05  & 1.0 & 40000 & 7.5 & 1 & 1. 50 & 1 & 1.0 & 0 \\

\hline
\end{tabular}

\end{minipage}
\end{table*}

\subsection{Bipolar nebulae}
\label{bipolar.sec}

The main goal of this paper is to find a model capable of producing a large-scale bipolar nebula with small secondary lobes near the waist of the bipolar, similar to those observed in Hubble~5 and K3--17. Figures (\ref{model_9.fig}) through (\ref{model_15.fig}) show some of the simulations that we performed. These have been selected, because they all show one or more features that have actually been observed or that we consider of general interest. In most figures one particular view is shown of the density integrated along the line of sight together with a long-slit position velocity diagram (P-V diagram). To obtain the P-V diagrams a simulated spectroscopic slit was placed along the main axis of the dense shell, which in most views coincides with the direction of the bipolar. For best appreciation of the secondary lobes, in the images and P-V diagrams the stationary outer regions of the low-density AGB-wind has been filtered out before the rendering. The specific parameters of each model are listed in Table \ref{bipolars.tab}.

\begin{figure}
\includegraphics[width=40mm]{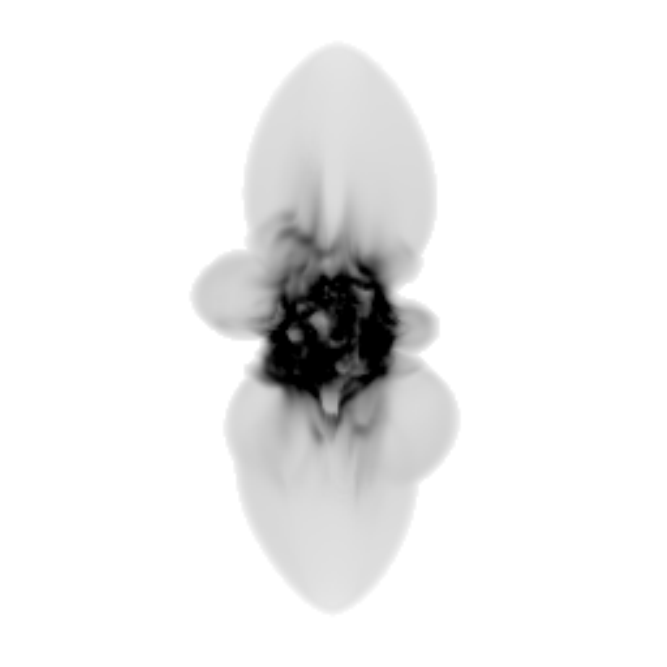}
\includegraphics[width=40mm]{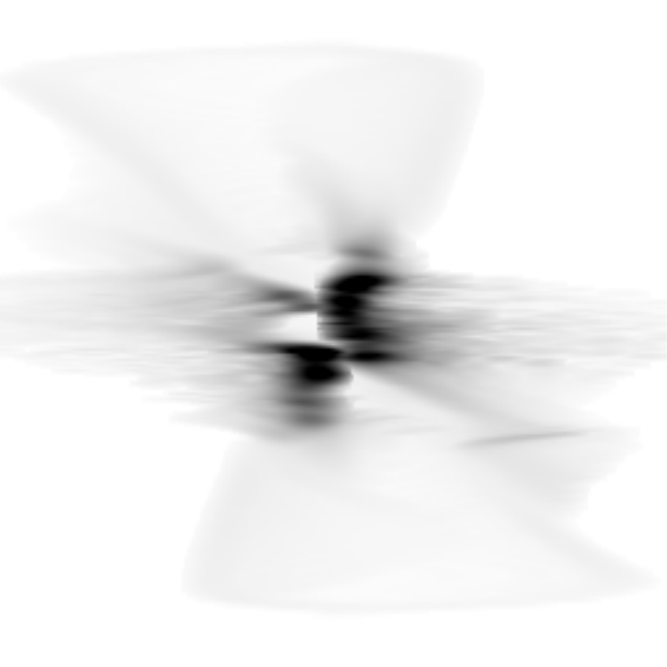}

\caption{The image and position-velocity diagram of Model B9 are shown. The inverted image intensity corresponds to the integral of the density squared along the line of sight, with a power-law gray-scale with an exponent of 0.25.}
\label{model_9.fig}
\end{figure}

\begin{figure}
\includegraphics[width=40mm]{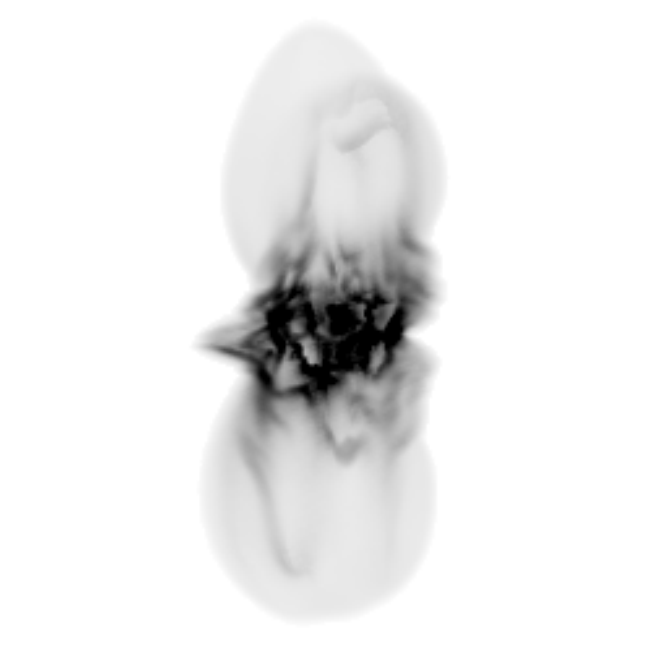}
\includegraphics[width=40mm]{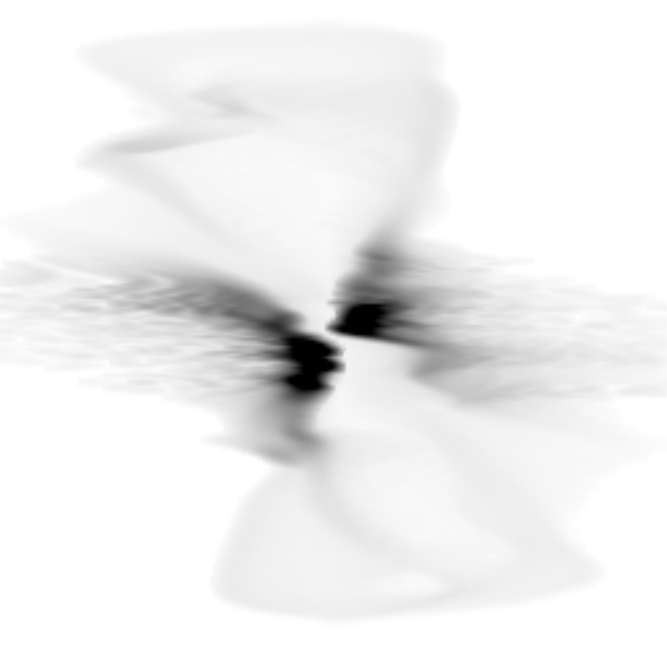}
\caption{Model B6. The gray-scale is the same as that in the previous figures.}
\label{model_6.fig}
\end{figure}

\begin{figure}
\includegraphics[width=80mm]{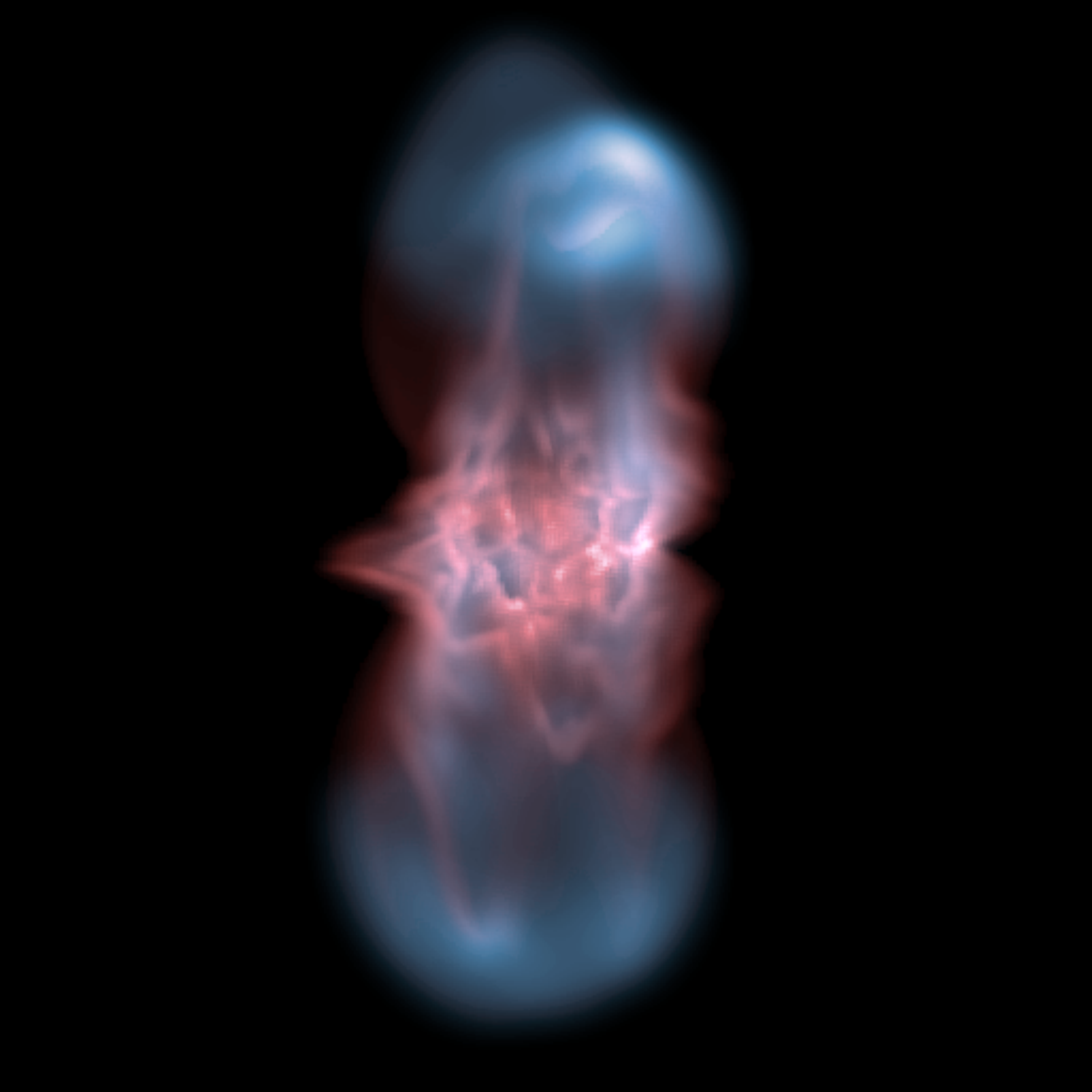}
\caption{For Model B2 the emission measure for cool high density gas is color coded in red and the diffuse free-free emission as a measure for the expected X-ray emission is coded in blue. The X-ray emission has been smoothed with a gaussian kernel to simulate the lower spatial resolution of X-ray observations.}
\label{model_6_ff.fig}
\end{figure}

\begin{figure}
\includegraphics[width=40mm]{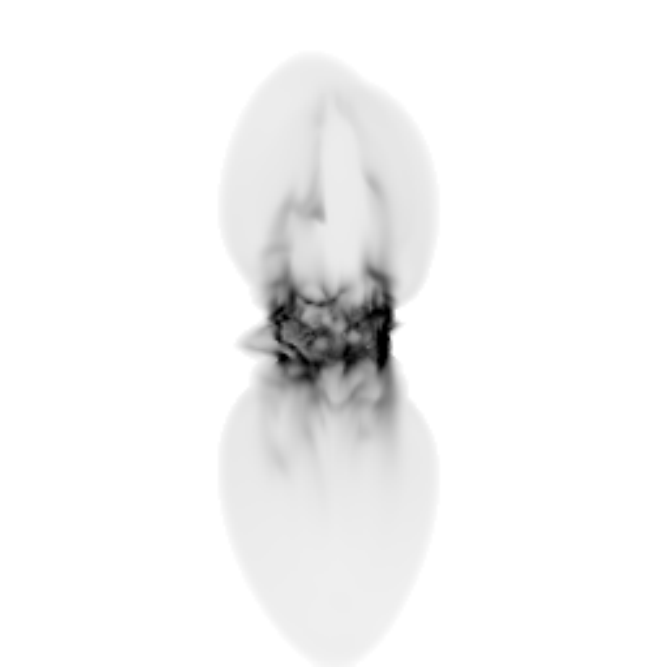}
\includegraphics[width=40mm]{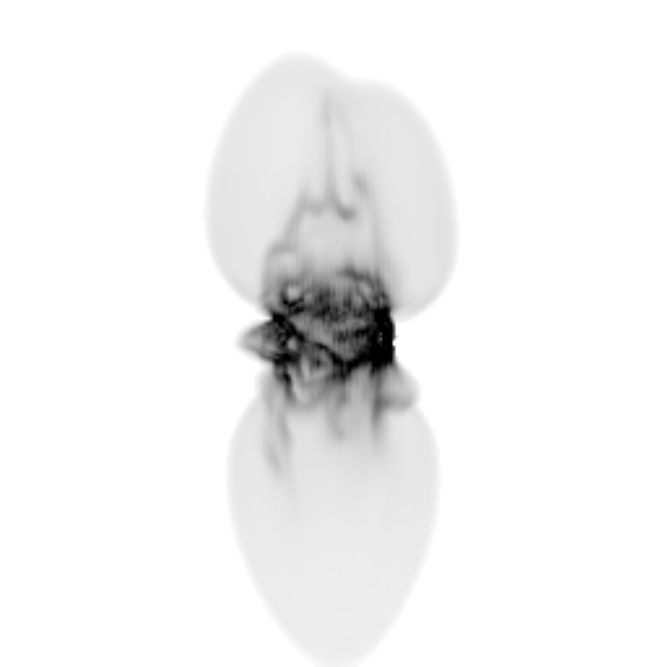}
\caption{ Two different views of Model B2. In the second image the view has been rotated around the objects axis by 35 degrees. The gray-scale is the same as that in the previous figures.}
\label{model_2.fig}
\end{figure}

\begin{figure}
\includegraphics[width=40mm]{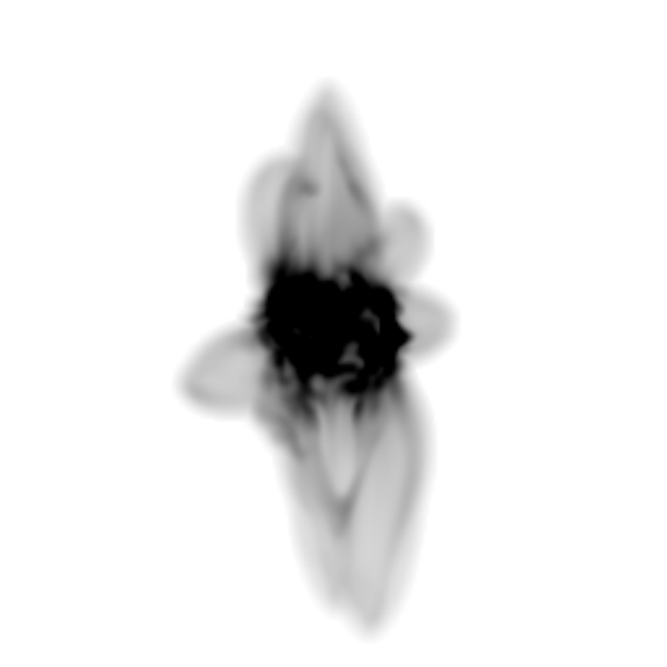}
\includegraphics[width=40mm]{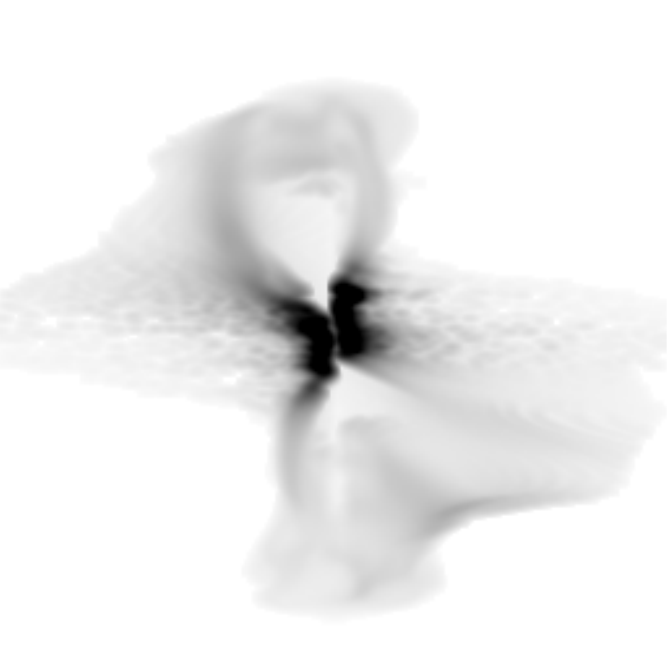}
\caption{Image and P--V diagram of Model B14. The gray-scale is the same as that in the previous figures.}
\label{model_14.fig}
\end{figure} 	

\begin{figure}
\includegraphics[width=80mm]{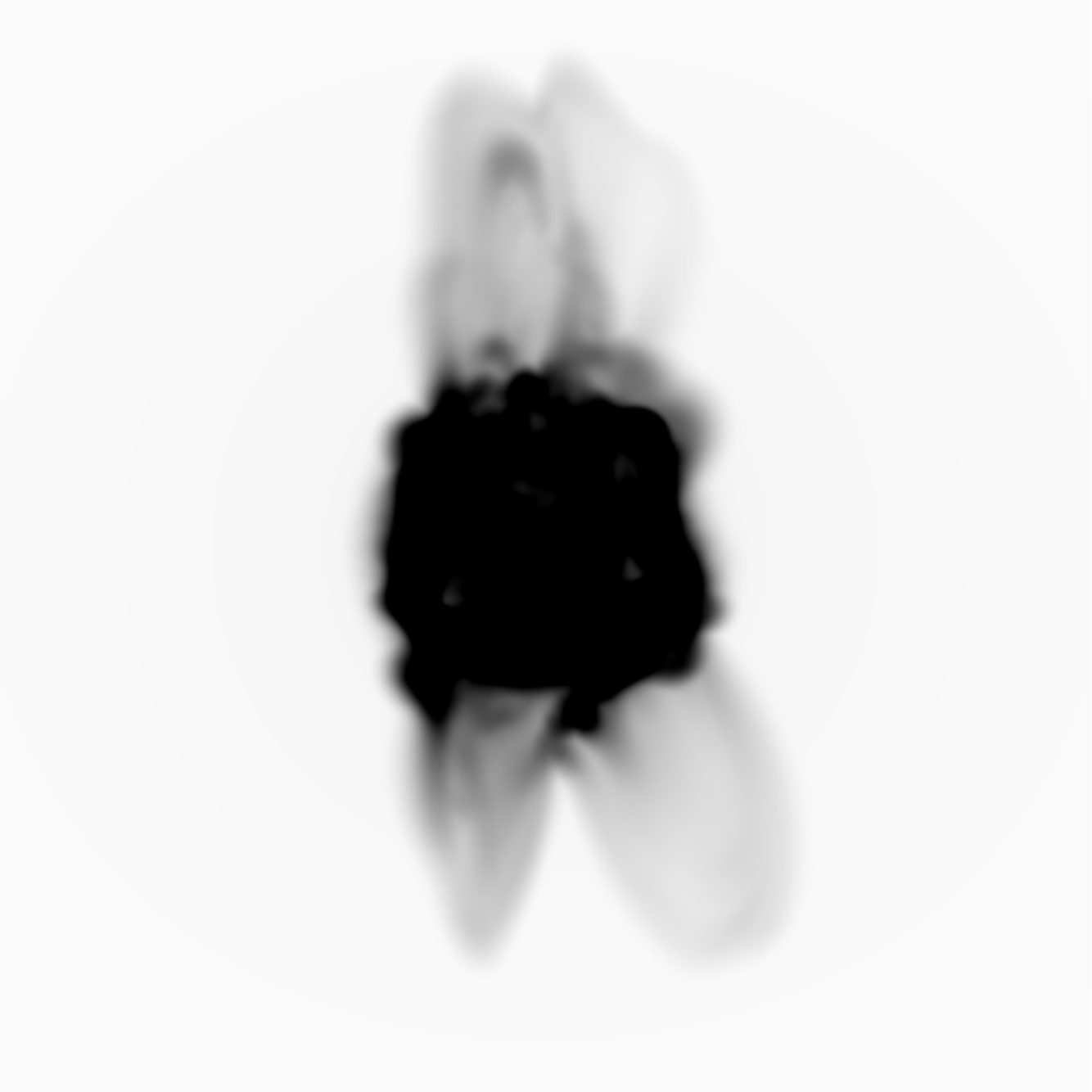}
\caption{Image of Model B15. The gray-scale is the same as that in the previous figures.}
\label{model_15.fig}
\end{figure} 	

The bipolar models have been computed to a size that is still about a factor of approximately three smaller than the actual size of Hubble~5 and represent the initial stage of the evolution. This was necessary to resolve the filamentary structure with the 256$^3$ grid used in all the bipolar models.

The similarity between the observations and the simulations in terms of structure and shape of the position-velocity diagram is striking. See, for example, Model B9 in Figure (\ref{model_9.fig}). The bipolar bubble structure with the secondary bubbles is very similar to that of Hubble~5 (Figure \ref{observations.fig}). The P-V diagram clearly shows that expansion of the nebula is non-homologous, since the outline of the spectrum is very different from that of the image.
Model B9 also shows a very complex array of secondary lobes. Several of these have merged with the main lobes, similar to what was found for Hubble~5 in the morpho-kinematic model by L\'opez et al. (2012).

Model B6 in Figure (\ref{model_6.fig}) is noteworthy because of its similarity to Hubble~5 and K3-17 in terms of its slight point-symmetry and small-scale features. The point-symmetry is due the an inclination of the fast wind axis with respect to the axis of symmetry of the dense shell. The effect of this tilt is, however, much less than expected, with only a light S-shaped distortion or point-symmetric bumps in bipolar lobes, which otherwise continued to be well aligned with the axis of the dense shell. The reason of the relatively small distortion of the bipolar structure is that the collimated outflow is readily redirected towards the main axis by the interaction with the denser regions of the shell.

The high density ``streamers'' in the P-V diagram that emerge from the high density shell, are due to the acceleration of the high density filaments. Similar streamers have been observed in Hubble~5 (L\'opez et al., 2012).

In Figure (\ref{model_6_ff.fig}) we have rendered the expected free-free emission of the gas with temperature above $10^6$~Kelvin as a measure for the expected X-ray emission. This emission is shaded in blue, whereas the cool dense gas is shaded in red. Note that, as opposed to the case of the starfish models (Figure \ref{starfish_xray.fig}) in the bipolar model the X-ray emission is strongest in the bow-shock regions of the main lobes and is likely to dominate with respect to the central region.

Model B2 in Figure (\ref{model_2.fig}) is notable because of its apparent multiple outflows inside and outside of the main bipolar. This is due to the higher density contrast between the filaments and the voids in the dense shell. The view shown in the right panel has similarities to the structure of NGC~7026 (Clark et al., 2012), suggesting that the origin of the multipolar appearance of this object might be found in the interaction of the fast stellar wind with a dense AGB-shell with a complex structure. NGC~7026 does not show small secondary lobes that would be as conspicuous as those in Hubble~5. This might indicate that in NGC~7026 the density contrast between filaments and voids is not high enough in the equatorial region.

Model B14 in Figure (\ref{model_14.fig}) shows a very complex array of secondary lobes. The fast wind velocity is this model is lower than in the others and the cooling of the bipolar shock wave is very strong. The wind is strongly collimated by the dense shell and the main bipolar lobes do not expand much laterally and develop a more jet-like appearance.

Finally, Model B15 in Figure (\ref{model_15.fig}) has a particularly high mass-loss rate resulting in a denser ambient medium and strongly collimated jet-like fingers very reminiscent to those in CRL~618 and similar objects. Most notably, in this model the fast wind is completely spherically symmetric. The collimation of the jets is entirely due to the interaction with the dense shell and its void structures. Hence, no mechanism for jet formation near the central star or even precession or intermittency of ejection in different directions is required as has been proposed (e.g. Vel\'azquez et al., 2011; Lee \& Sahai, 2003).

\section{Discussion}
\label{discussion.sec}

We performed hydrodynamical simulations to explore a possible formation mechanism of multiple lobes in planetary nebulae. In particular we investigate nebulae that show a large scale bipolar structure with small secondary lobes, such as those in Hubble~5 and K3--17. The proposed formation mechanism is based on the expansion of a fast low-density wind through a filamentary high-density shell at some distance to the central star.

We find that an inhomogeneous shell around a post-AGB star with a low-density fast wind can generate
multipolar nebulae if the structure of the shell consists of high density filaments with voids in between.
The density contrast between the filaments, the voids and the fast wind has to be such that the expansion speed
of the filaments remains much lower than that of the voids after being shocked and accelerated by the wind.
Filaments with lowest density may be accelerated to form part of the bipolar lobes or expand at an intermediate
speed to form filaments within the main lobes engulfed and compressed by the fast wind. Such medium density filaments can produce strong deformations of the bipolar lobes.

The result of the wind-shell interaction is a starfish nebula if the spherical dense shell and the fast wind have no systematic change of density and velocity as a function of angular distance from the equator. The size and expansion
speed of the individual lobes varies, depending on the details of the
void density distribution in the dense shell. In cases where the void structure approaches the size of the shell, individual lobes may be significantly larger than the others and produce a structure dominated by a single lobe.

In order to generate the bipolar structures similar to those of Hubble~5 and K3--17 it is not necessary
that the fast wind be collimated, only a poloidal density gradient in the dense shell is required. However, to produce a bipolar of a shape very similar to those observed in Hubble~5 and K3--17 with their secondary lobes, we always needed a wind with a velocity gradient from the equator to the poles. Furthermore, to produce the necessary shape of the lobes and at the same time the secondary lobes, the lobe expansion must be sufficiently fast such that the cooling time is of the order of the expansion time. This smoothes the filamentary structure and reduces the irregular expansion of the lobes. To achieve this, the wind has to be very fast and rather dense or the mass-loss rates smaller than what would be expected for the late AGB. Otherwise the strong cooling produces bipolar structures that are much more collimated than the lobes of Hubble~5 and K3--17. This is consistent with a high mass progenitor star.

The influence of cooling is very strong on the overall structure, since the shell with a density gradient produces
strong collimation of the fast wind. If the velocity of the fast wind is such that the expanding shock is strongly cooling, then the bipolar structure will be very narrow, similar to CRL 618. In order to generate wide lobes as in Hubble~5 or K3-17, the initial expansion must be adiabatic and the cooling must occur on the timescale of the nebula's lifetime. This is, however, only possible if the fast wind drastically reduces its power shortly after the initial interaction with the dense shell. Otherwise the adiabatic expansion will continue, since the density outside the shell probably drops faster than $r^{-2}$. We therefore conclude that the most powerful phase of the fast wind persists only for times of a hundred years or so. This, again, might be an indication for a massive central star (Sch\"onberner \& Steffen, 2007), consistent with the 5~~M$_{\odot}$ estimated by Pottasch \& Surendiranath (2007).

If the average size scale of the voids in the dense shell is smaller than the shell thickness, even though there are more voids, the number of lobes in a starfish nebula may become smaller. The smaller lobes become suppressed, since they may have to percolate through several voids to finally emerge. As the fast wind percolates through the small voids a warm and diffuse halo forms around the cold dense shell.

In L\'opez et al. (2012) we presented a 3D morpho-kinematic model for Hubble~5 based on a detailed data set of long-slit spectra and HST imaging. The model showed several results that where unexpected from a visual inspection of the data. First, the bipolar structure has large point-symmetric bumps along the line of sight, such that they are not visible in the image. These bumps have been attributed to an oblique bipolar collimated outflow that interacts with the main lobes at the same time generating the high brightness ``S''-structure that flows along the main lobes. It is, however, not immediately obvious whether the bipolar lobes and the point-symmetric bump can be formed from a single outflow that is inclined with respect to the equator of the high density shell or whether the main lobes and the bumps where formed by separate outflow events with differing directions. This work shows that such point-symmetric bumps can be produced with the same inclined bipolar outflow. No second highly collimated jet is necessary.

Second, in the morpho-kinematic model there are at least two, probably more, secondary lobes that have merged with the main lobes, such that they are visible only as bumps or silhouettes on the main lobes. The model also has secondary ``half" lobes sitting on the main lobes. Similar structures have been found in our hydrodynamic simulations, confirming the interpretation of these structures as secondary lobes. They are best seen in Figure (\ref{model_9.fig}). This shows that the multiplicity of the secondary lobes is likely even higher than that derived from those that are clearly separate from the main lobes. A detailed study of the secondary lobes and a high resolution imaging and spectroscopic study of the dense shell might provide information about the ejection process and, possibly, about the atmospheric turbulence in the stellar atmosphere before the ejection of the dense shell.

\section{Concluding remarks}
\label{conclusion.sec}

The main conclusion from this work is that secondary lobes in planetary nebulae, such as Hubble~5 and K3--17, can be formed through the interaction of a fast low-density wind with a complex high density environment, most likely a filamentary circumstellar shell. This provides a natural and simple alternative scenario that does not require intermittent precessing collimated outflows. The results do also suggest that the progenitor of the central star was rather massive, above about 5~$M_{\odot}$.

{\bf Acknowledgements} \\
We thank Alejandro Raga for his support during the development of the hydrodynamics module of Shape. This work has been supported by {\em Universidad Nacional Aut\'onoma de M\'exico} through
grant DGAPA-PAPIIT IN 100410. W.S. acknowledges support by the {\em Alexander von Humboldt Foundation} through the F.-W. Bessel Award and is very thankful for the hospitality of the {\em Technische Universit\"at Braunschweig}, Germany, during his sabbatical year. N.K. acknowledges support from the Killam Trusts and travel support from the {\em Technische Universit\"at Braunschweig}.

\bibliographystyle{mnras}

\begin{thebibliography}{99}

\bibitem[Balick \& Frank, 2002]{BF02}
Balick, B., Frank, A., 2002, Annu. Rev. Astron. Astrophys., 40, 439–86

\bibitem[Bujarrabal et al., 1998]{BAN98}
Bujarrabal, V., Alcolea, J., Neri, R., 1998, ApJ, 504, 915

\bibitem[Chesneau et al., 2006]{Chesneau06}
Chesneau, O., Collioud, A., De Marco, O., Wolf, S., Lagadec, E., Zijlstra, A. A., Rothkopf, A., Acker, A., Clayton, G. C., L\'opez, B., 2006, A\&A, 455, 1009-1018

\bibitem[Chong et al., 2012]{Chong12}
Chong, S.-N., Kwok, S., Imai, H., Tafoya, D., Chibueze, J., 2012,
ApJ, 760, 115

\bibitem[Clark et al., 2012]{Clark12}
Clark, D.M., L\'opez, J.A., Steffen, W., Richer, M.G., 2012, ApJ, 145, 57

\bibitem[Cox et al., 2012]{Cox12}
Cox, N. L. J., Kerschbaum, F., van Marle, A.-J., et al., 2012, A\&A, 537, 35

\bibitem[Cox et al., 2000]{Cox00}
Cox, P., Lucas, R., Huggins, P.J., Forveille, T., Bachiller, R., Guilloteau, S., 2000, A\&A, L25-L28

\bibitem[Courant et al. 1967]{CFL}
Courant, R., Friedrichs, K, \& Lewy, H., IBM J. Res. Dev.. 11, 215

\bibitem[Dalgarno \& McCray, 1972]{DM72}
Dalgarno, A., McCray, R.A., 1972,
ARA\&A, 10, 375

\bibitem[Freytag \& H\"ofner, 2008]{FH08}
Freytag, B., H\"ofner, S., 2008, A\&A, 483, 571-583

\bibitem[Garc\'ia-D\'iaz et al., 2012] {GD12}
Garc\'ia-D\'iaz, Ma. T., L\'opez, J.A., Steffen, W., Richer, M.G., ApJ, 761, 172

\bibitem[Garc\'ia-Segura, 2010] {GS10}
Garc\'ia-Segura, G., A\&A, 520, id. L5, 4pp.

\bibitem[Izumiura et al., 1996]{I96}
Izumiura, H., Hashimoto, O., Kawara, K., Yamamura, I., Waters, L. B. F. M., 1996,
A\&A, 315, L221-L224

\bibitem[Kahn \& West, 1985]{KW85}
Kahn, F.D., West, K.A., 1985, MNRAS, 212, 837-850

\bibitem[Kwok, Purton \& Fitzgerald]{KPF78}
Kwok, S., Purton, C. R., \& Fitzgerald, P. M., 1978, ApJ, 219, L125-L127

\bibitem[Lee \& Sahai, 2003]{LS03}
Lee,  C.-F., Sahai, R., 2003, ApJ, 586, 319

\bibitem[Lewis, 1989]{Lewis89}
Lewis, J.P., 1989, ACM SIGGRAPH {\em Computer Graphics}, Vol. 23, Num. 3, 263-270

\bibitem[L\'opez et al., 2012]{Lopez12}
L\'opez, J. A. , Garc\'ia-D\'iaz, Ma. T., Steffen, W., Riesgo, H., Richer, M.G., 2012, ApJ, 750, 131

\bibitem[L\`opez et al., 1998]{Lopez98}
L\'opez, J.A., Meaburn, J., Bryce, M., Holloway, A.J., 1998, ApJ, 493, 803-810

\bibitem[L\`opez et al., 1993]{LMP93}
L\'opez, J.A., Meaburn, J., Palmer, J.W., 1993, ApJ, 415, L135-L137

\bibitem[L\`opez et al., 1995]{LVR95}
L\'opez, J.A., V\'azquez, R., Rodr\'{\i}guez, L.F., 1995, ApJ, 455, L63-L66

\bibitem[Manchado, Stanghellini \& Guerrero, 1996]{MSG96}
Manchado, A., Stanghellini, L., Guerrero, M.A., 1996, ApJ, 466, L95-L98

\bibitem[Perlin 2002]{Perlin02}
Perlin, K., 2002, SIGGRAPH '02 Proceedings of the 29th annual conference on computer graphics and interactive techniques,
vol. 21, issue 3, pages 681-682

\bibitem[Pottasch \& Surendiranath, 2007]{PS07}
Pottasch, S. R.; Surendiranath, R., 2007, A\&A, 462, 179-192

\bibitem[\protect\citeauthoryear{Raga et al.}{2000}]{Raga00}
Raga, A.C., Navarro-Gonz\'alez R., Villagr\'an-Muniz, M., 2000, RevMexAA, 36, 67-76

\bibitem[Ribeiro et al., 2011]{ribeiro11}
Ribeiro, V.A.R.M., Darnely, M.J., Bode, M.F., Munari, U., Harman, D.J., Steele, I.A., Meaburn, J.,
2011, MRNAS, 412, 1701-1709

\bibitem[Sahai, 2000]{Sahai00}
Sahai, R., 2000, ApJ, 537, L43-L47

\bibitem[Sahai \& Trauger, 1998]{ST98}
Sahai, R., Trauger, J.T., 1998, ApJ, 116, 1357-1366

\bibitem[S\'anchez Conteras \& Sahai, 2001]{SS01}
S\'anchez Conteras, C., Sahai, R., 2001, ApJ, L173-L176

\bibitem[Santander--Garc\'{\i}a et al, 2012]{S12}
Santander-Garc\'{i}a , M., Bujarrabal, V., Alcolea, J., 2012,
A\&A, 545, 114

\bibitem[Sch\"onberner \& Steffen]{SS07}
Sch\"onberner, D., Steffen, M., 2007, ASP Conference Series, eds. Kerschbaum, Charbonnel \& Wind,
Vol., 378, 343-344

\bibitem[\protect\citeauthoryear{Steffen}{2009}]{Stef2009b}
Steffen W., Garc\'{i}a--Segura G., Koning N., 2009b, ApJ, 691, 696

\bibitem[Steffen, et al.(2011)]{skwmm11}
Steffen, W., Koning, N., Wenger, S., Morisset, C., Magnor, M., 2011, ''Shape: A 3-D Modeling Tool for Astrophysics'', IEEE Trans. on Visualization \& Computer Graphics, vol. 17, no. 4, pp. 454-465 (arXiv:1003.2012).

\bibitem[\protect\citeauthoryear{Steffen \& L\'opez}{2004}]{SL04}
Steffen W., L\'{o}pez J. A., 2004, ApJ, 612, 319

\bibitem[\protect\citeauthoryear{Steffen}{2006}]{Stef2006}
Steffen W., L\'{o}pez J. A., 2006, RevMexAA, 42, 99

\bibitem[Trammel \& Goodrich]{TG02}
Trammel, S.R., Goodrich, R.W., 2002, ApJ, 668-693

\bibitem[van Leer, 1974]{vl74}
Van Leer, B., 1974, J. Comput. Phys., 14, 263

\bibitem[van Leer, 1982]{vL82}
Van Leer, B., 1982, ICASE Report No. 82-30

\bibitem[Vel\'azquez et al., 2012]{V12}
Vel\'azquez, P.F., Raga, A.C., Riera, A., Steffen, W., Esquivel, A., Cantó, J., Haro-Corzo, S., 2012,
MNRAS, 419, 3529-3536

\bibitem[Vishniac, 1983]{V83}
Vishniac, E.T., 1983, ApJ, 274, 152

\bibitem[Wenger, et al.(2012)]{wenger12}
Wenger, S., Ament, M., Steffen, W., Koning, N., Weiskopf, D., Magnor, M., 2012, IEEE Computing in Science and Engineering, vol. 14, no. 3, pp. 78-87 (arXiv:1204.6132)

\end{thebibliography}

\end{document}